# Flux Loop Measurements of the Magnetic Flux Density in the CMS Magnet Yoke

V. I. Klyukhin

*Skobeltsyn Institute of Nuclear Physics, Lomonosov Moscow State University, RU-119992, Moscow, Russia*

Phone : +41-22-767-6561, fax : +41-22-767-7920, e-mail : Vyacheslav.Klyukhin@cern.ch

N. Amapane

*INFN Torino and the University of Torino, I-10125, Torino, Italy*

A. Ball, B. Curé, A. Gaddi, H. Gerwig, M. Mulders

*CERN, CH-1211, Geneva 23, Switzerland*

A. Hervé, R. Loveless

*University of Wisconsin, WI 53706, Madison, USA*

**Abstract** The Compact Muon Solenoid (CMS) is a general purpose detector, designed to run at the highest luminosity at the CERN Large Hadron Collider (LHC). Its distinctive features include a 4 T superconducting solenoid with 6-m-diameter by 12.5-m-length free bore, enclosed inside a 10,000-ton return yoke made of construction steel. The return yoke consists of five dodecagonal three-layered barrel wheels and four end-cap disks at each end comprised of steel blocks up to 620 mm thick, which serve as the absorber plates of the muon detection system. To measure the field in and around the steel, a system of 22 flux loops and 82 3-D Hall sensors is installed on the return yoke blocks. A TOSCA 3-D model of the CMS magnet is developed to describe the magnetic field everywhere outside the tracking volume measured with the field-mapping machine. The first attempt is made to measure the magnetic flux density in the steel blocks of the CMS magnet yoke using the standard magnet discharge with the current ramp down speed of 1.5 A/s.

## 1 Introduction

The muon system of the CMS detector includes a 10,000-ton yoke comprised of the construction steel plates up to 620 mm thick, which return the flux of the 4 T superconducting solenoid and serve as the absorber plates of the muon detection system [1–4]. During the LHC long shutdown occurring in 2013/14 the CMS magnet yoke was upgraded with the additional 14 m diameter end-cap disks at the extremes of the muon detection system [5].

The magnetic flux density in the central part of the CMS detector, where the tracker and electromagnetic calorimeter are located, was measured with precision of $7 \cdot 10^{-4}$ with the field-mapping machine at five central field values of 2, 3, 3.5, 3.8, and 4 T [6]. To describe the magnetic flux everywhere outside the measured volume, a three-dimensional (3-D) magnetic field model of the CMS magnet has been developed [7] and calculated with TOSCA [8]. The model reproduces the magnetic flux density distribution measured inside the CMS coil with the field-mapping machine within 0.1% [9]. The modification of this model for the upgraded CMS magnet yoke was validated by comparing the calculated magnetic flux density with the measured one in the selected regions of the CMS magnetic system. [10]

A direct measurement of the magnetic flux density in the yoke selected regions was provided during the CMS magnet test of 2006 with 22 flux loops of 315÷450 turns wound around the yoke blocks. The "fast" (190 s time-constant) discharges of the CMS coil made possible by the protection system, which is provided to protect the magnet in the event of major faults [11–12], induced in the flux loops the voltages caused by the magnetic flux changes. An integration technique [13–14] was developed to reconstruct the average initial magnetic flux density in steel blocks at the full magnet excitation, and the contribution of the eddy currents was calculated with ELECTRA [15] and estimated on the level of a few per cent [16].

The results of the magnetic flux measurements done with the flux loops and comparison the obtained values with the calculations performed with the previous TOSCA CMS magnet model are described elsewhere [17].

In present paper we compare the calculations based on the recent CMS magnet model with the measurements performed with the flux loops in the yoke steel and with the 3-D Hall probes installed at the steel-air interfaces in the gaps between the CMS yoke parts. The read-out system of the flux loop induced voltages is substantially improved during the LHC long shutdown and allows now to measure the small voltages of the order of 20 mV with the sampling rate of 4 Hz.

## 2 The CMS Magnet Model Description

The CMS magnet model for the upgraded detector is presented in Fig. 1. The central part of the model comprises the coil four layers of the superconductor and the full magnet yoke that consists of five barrel wheels of the 13.99 m inscribed outer diameter and 2.536 m width, two nose disks of 5.26 m diameter on each side of the coil, three large end-cap disks of the 13.91 m inscribed outer diameter on each side of the magnet, two small end-cap disks of 5 m diameter, and two additional 0.125 m thick end-cap disks of the 13.91 m inscribed outer diameter mounted around these small disks.

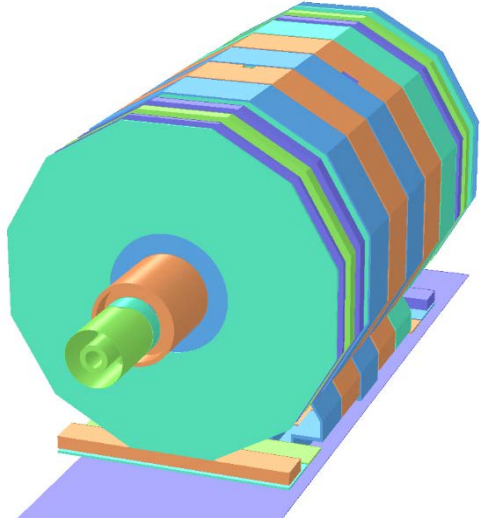

**Fig. 1** 3-D model of the CMS magnetic system

The total length of the central part is 21.61 m. On both sides of the central part the hadronic forward calorimeter absorbers and shields, the steel collars and rotating shields of the beam pipe are included into the model. The 40 mm thick steel floor of the experimental underground cavern has the length of 36.4 m and the width of 9.9 m.

Each barrel wheel except of central one has three layers of steel connected with brackets. The central barrel wheel comprises the fourth most inner layer, tail catcher, made of steel and turned by 5 degrees in the azimuth angle with respect to dodecagonal shape of the barrel wheels. The coordinate system used in the model corresponds to the CMS reference system where the X-axis is directed in horizontal plane toward the LHC center, the Y-axis is upward, and the Z-axis coincides with the superconducting coil axis and has the same direction as the positive axial component of the magnetic flux density.

The model comprises 21 conductors, the barrel feet and the end-cap disk carts and contains 7,111,713 nodes of the finite element mesh.

The dimensions of the yoke parts and the superconducting coil modules are described elsewhere [17]. The operational current of the CMS superconducting coil is 18.164 kA.

# 3 Comparison of the Measured and Calculated Magnetic Flux Density

The measurements used for the comparisons were obtained in the CMS magnet ramp down from 3.81 to 1 T central field with the current ramp down speed of 1.5 A/s and a subsequent fast dump of the current to zero on July 17–18, 2015. The measurements of the voltages with maximum amplitudes of 20–200 mV induced in the flux loops were performed during 15000–17000 s. The flux loops measurements are complemented with measuring the magnetic flux density with the 3-D Hall sensors installed between the barrel wheels and on the first end-cap disks at Z-coordinates of 1.273, –1.418, –3.964, –4.079, –6.625, and –7.251 m. The sensors are aligned in the rows at the Y-coordinates of –3.958, –4.805, –5.66, and –6.685 [10]. The average magnetic flux density in steel was reconstructed by the off-line integration of the voltages [16] as shown in Fig. 2.

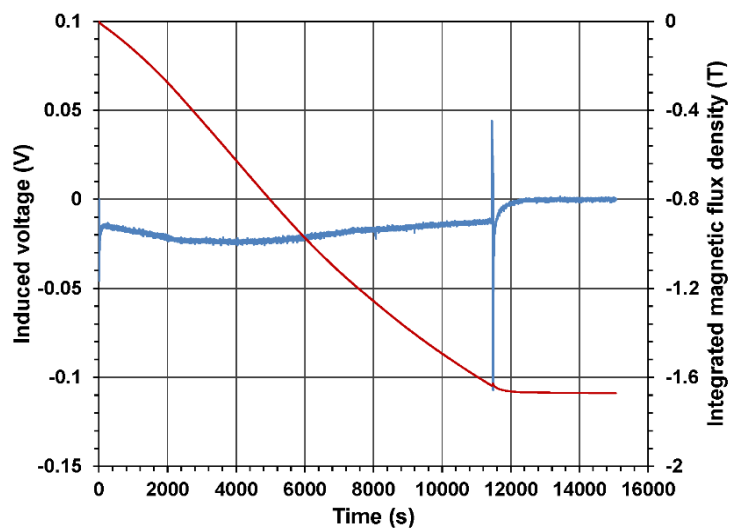

**Fig. 2** Induced voltage (left scale, noisy curve) and the integrated average magnetic flux density (right scale, smooth curve) in the tail catcher block cross-section at Z=0 m and at the operational current of 18.164 kA. The rapid maximum and minimum voltage at 11445 s corresponds to the transition from the standard discharge to the fast discharge of the magnet

In Figs. 3–4 the measured values of the magnetic flux density vs. Z- and Y-coordinates are displayed and compared with the calculated field values obtained with the CMS magnetic system model at the operational current of 18.164 kA.

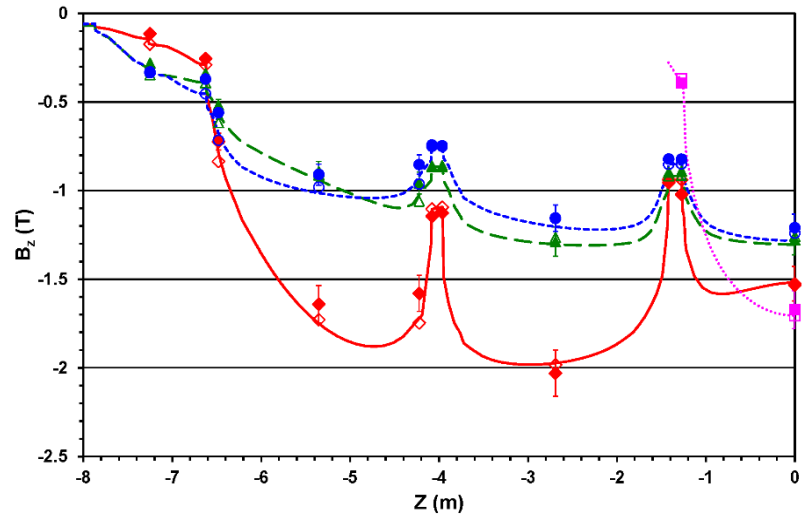

**Fig. 3** Axial magnetic flux density measured (filled markers) and calculated (opened markers) in the tail catcher (squares), first (rhombs), second (triangles), and third (circles) barrel layers vs. Z-coordinate at the operational current of 18.164 kA. The lines represent the calculated values along the Hall sensors locations at the positive X-coordinates and at the Y-coordinates of –3.958 (small dotted line), –4.805 (solid line), –5.66 (dashed line), and –6.685 (dotted line) m, corresponded to the tail catcher, first, second, and third barrel layer blocks

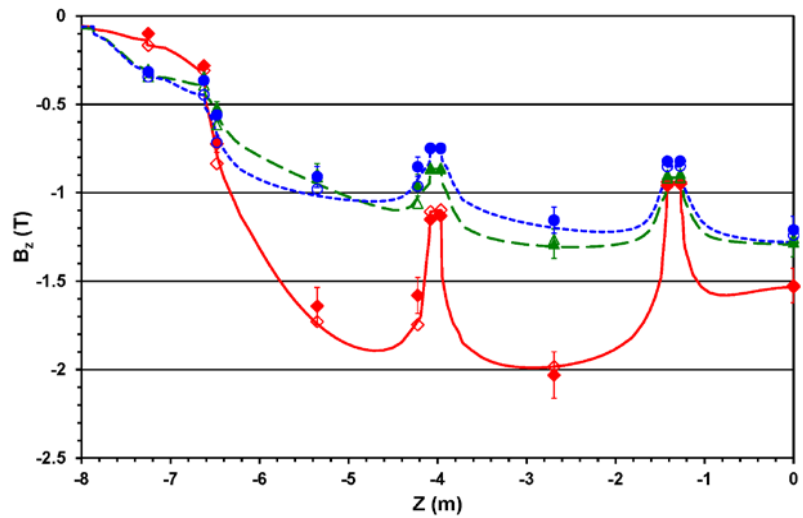

**Fig. 4** Axial magnetic flux density measured (filled markers) and calculated (opened markers) in the first (rhombs), second (triangles), and third (circles) barrel layers vs. Z-coordinate at the operational current of 18.164 kA. The lines represent the calculated values along the Hall sensors locations at the negative X-coordinates and at the Y-coordinates of –4.805 (solid line), –5.66 (dashed line), and –6.685 (dotted line) m, corresponded to the first, second, and third barrel layer blocks

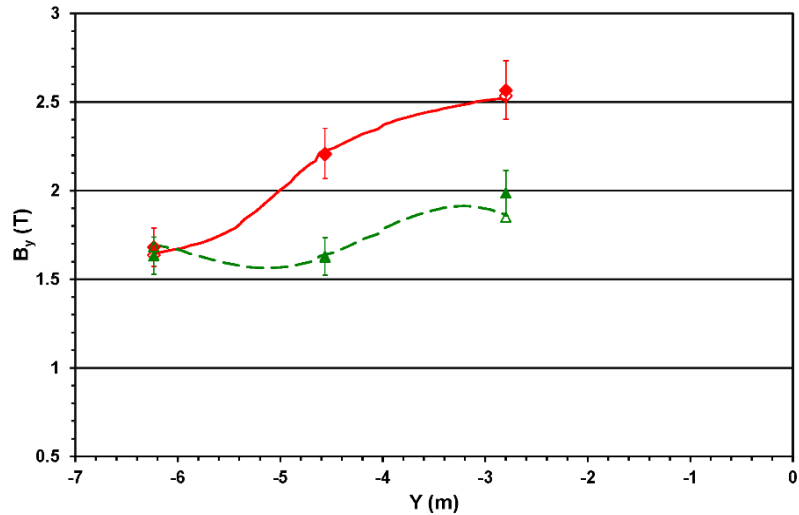

**Fig. 5** Axial magnetic flux density measured (filled markers) and calculated (opened markers) in the first (rhombs) and second (triangles) end-cap disks vs. Y-coordinate at the operational current of 18.164 kA. The lines represent the calculated values in the middle planes of the end-cap disks

The comparison gives the differences between the calculated and measured values of the magnetic flux density in the flux loop cross-sections as follows: (6.96±8.6) % in the barrel wheels and (−1.37±3.29) % in the end-cap disks. The error bars of the magnetic flux density measured with the flux loops are of ±6.44% and aroused from the errors in the knowledge of the flux loops geometries. The difference between the calculated and measured magnetic flux density in the Hall sensors locations is (4.32±7.12) % at the negative X-coordinates and (3.77±9.0) % at the positive X-coordinates. The error bars of the 3-D Hall sensor measurements are ± (0.012±0.005) mT.

## 4 Conclusions

The first attempt of measuring the magnetic flux density in the steel blocks of the CMS magnet flux return yoke is made using the flux loop technique and the standard magnet discharge with the current ramp down speed of 1.5 A/s. Caused by very small maximum amplitudes of the voltages induced in the flux loops during the standard magnet discharge, the averaged precision of the present measurements is worse than the averaged accuracy of the measurements performed using the CMS magnet fast discharges [10], but still allows to confirm the correctness of the CMS magnetic flux description done with the CMS magnetic system model.